\documentclass[twocolumn,aps,superscriptaddress,showpacs,nofootinbib,floatfix]{revtex4-1}   	
\usepackage{graphicx}				
\usepackage{amsmath}								
\usepackage{amssymb}
\usepackage{soul}

\usepackage[normalem]{ulem}
\usepackage{multirow}
\usepackage{xcolor}

\renewcommand\sout{\bgroup \color{red} \ULdepth=-.5ex \ULset}

\begin{document}


\title{ $X(3872)$ and $T_{cc}$: structures and productions in heavy ion collisions}

\author{Hyeongock Yun}
\email{mero0819@yonsei.ac.kr}
\affiliation{Department of Physics and Institute of Physics and Applied Physics, Yonsei University, Seoul 03722, Korea}
					
\author{Daeho Park}
\email{olmu100@yonsei.ac.kr}
\affiliation{Department of Physics and Institute of Physics and Applied Physics, Yonsei University, Seoul 03722, Korea}
\author{Sungsik Noh}
\email{sungsiknoh@yonsei.ac.kr}
\affiliation{Department of Physics and Institute of Physics and Applied Physics, Yonsei University, Seoul 03722, Korea}				
		
\author{Aaron Park}
\email{aaron.park@yonsei.ac.kr}
\affiliation{Department of Physics and Institute of Physics and Applied Physics, Yonsei University, Seoul 03722, Korea}

\author{Woosung Park}
\affiliation{Department of Physics and Institute of Physics and Applied Physics, Yonsei University, Seoul 03722, Korea}

\author{Sungtae Cho}
\email{sungtae.cho@kangwon.ac.kr}
\affiliation{Division of Science Education, Kangwon National University, Chuncheon 24341, Korea}

\author{Juhee Hong}
\affiliation{Department of Physics and Institute of Physics and Applied Physics, Yonsei University,
Seoul 03722, Korea}

\author{Yongsun Kim}%
\email{kimy@cern.ch}
\affiliation{Department of Physics, Sejong University, Seoul, Korea  }

\author{Sanghoon Lim}%
\email{shlim@pusan.ac.kr}
\affiliation{Department of Physics, Pusan National University, Pusan, Korea  }

\author{Su Houng Lee}%
\email{suhoung@yonsei.ac.kr}
\affiliation{Department of Physics and Institute of Physics and Applied Physics, Yonsei University, Seoul 03722, Korea}

\begin{abstract}
We argue why the recently observed $T_{cc}$ could either be  a  compact multiquark configuration  or a loosely bound molecular configuration composed of charmed mesons, whereas the $X(3872)$ 
is most likely a molecular configuration.  The argument is based on  different
short range interactions for these tetraquark states coming from the color-color and color-spin interaction in a quark model, and the presence of a common strong D-wave mixing at larger distance similar to the deuteron case, which for the molecular configurations lead to large sizes. 
Such an analogy at large distance allows us to calculate the transverse momentum dependence of the loosely bound molecular configuration of tetraquarks produced in heavy ion collisions using the coalescence model that successfully reproduces the deutron data using the proton spectra. 
The ratio of the integrated $X(3872)$ yield obtained from our method to the $\psi(2S)$ yield obtained from statistical hadronization model method is calculated to be $0.806 \pm 0.234$, which is a factor of 2.47 larger than that obtained by using statistical model predictions for both particles and in line with the data from the CMS experiment. 
As the previously calculated  transverse momentum distribution of the $T_{cc}$ assuming the structure to be a compact multiquark configuration is markedly different, experimental measurements of the transverse distribution of the tetraquark states will discriminate between their two possible structures.  
\end{abstract}

\maketitle

\section{Introduction}

Besides the historical discovery of Higgs boson, one of the greatest victories of Large Hadron Collider (LHC) is the discovery of numerous new hadrons, more than 50 over the past decade~\cite{CERN-article2021}. In particular, observation of several resonances composed of four or five quarks (and anti-quarks) unambiguously confirmed the existence of tetraqurk and pentaquark states, so called exotic particles.   These results provide novel inputs for understanding the nature of strong force in the hadronic scale, thus providing crucial clue to comprehend the property of confinement\cite{Belle:2003nnu, LHCb:2015yax,LHCb:2019kea,D0:2016mwd,LHCb:2021vvq}.   The most important remaining question would be the internal structure; are they in molecular system composed of loosely bound hadrons, or are they tight multiquark states where the quarks and anti-quarks form compact configurations with the size of a typical hadron?  No clear answer has been reported so far. 
 
In this paper, we discuss $X(3872)$ ($c\bar{c}u\bar{u}$, a hidden-charm tetraquark) and $T_{cc}$ ($cc\bar{u}\bar{d}$, open-charm tetraquark), which are of extraordinary interest for their conjugate relation in terms of charm quark flavor. Furthermore, they are the representative candidates that could either be compact multiquark configurations or  loosely bound molecular configurations.  We provide strong arguments that $X(3872)$ must be in molecular state of $D \bar{D}^*$ based on two evidences; (a) one is the absence of strong short range attraction required for a compact configuration, and (b) the other is the presence of a strong D-wave mixing coming from pion exchange, a mechanism crucial for the binding of the deuteron. 
On the other hand, we report that the $T_{cc}$ has possibilities to be either a compact multiquark state or a loosely bound molecular state of $DD^*$ due to two competing effects: (a) the presence of a strong short range attraction, and (b)  a strong mixing of the D-wave configuration of $DD^*$ by the $\pi$-exchange that is similar but weaker than that present in $X(3872)$.

We also show that the transverse momentum distributions calculated for both the compact mutiquark and loosely bound molecular states are completely different, thus allowing the discrimination of their structures through high statistics measurements at the LHC heavy ion experiments in near future.   Such measurement will need data at lower transverse momentum than that reported in  
the recent anomalous enhancement of $X(3872)$ production in heavy ion collision experiments\cite{CMS:2021znk}. 

The paper is organized as follows.  In Sec. II, we discuss the short range interaction of tetraquark states based on a quark model.  In Sec. III, we discuss the D-wave mixing.  In Sec. IV, we discuss the transverse momentum distribution for the deuteron and $^3$He measured in heavy ion collision and the fit from a coalescence model. In Sec. V, we use the same coalescence model to predict the distribution for the molecular configuration of tetraquarks and compare the result assuming a compact configuration.  Finally, we give the summary in Sec. VI.  The appendix provides the fitting formula and details of the coalescence model used in this work.

\section{Short distance attractions}

In a previous publication\cite{Park:2019bsz}, we have shown that the short distance parts of
the baryon-baryon interactions for various quantum numbers from the recent lattice calculation can be well
reproduced using a constituent quark model with color-spin interaction. 
In fact, the full constituent quark model calculations are found to be comparable to the  simple estimates based on the following color spin factor for a multiquark configuration\cite{Park:2019bsz}.
\begin{align}
  K&=-\sum_{i<j}^n \frac{1}{m_im_j} \lambda^c_i \lambda^c_j  \sigma_i \cdot \sigma_j, \label{color-spin0} 
\end{align}
where $\lambda_i^c$ and $\sigma_i$ are, respectively, the color and spin operators and $m_i$ the mass of the $i$'th quark among $n$ quarks.   
Although a full quark model calculation will include the spatial wave function, if the interquark distancees among quarks are all the same, the short distance interaction between hadrons can be well described using only the color-spin factor times an overall factor that determines the interaction strength at the hadron size. 

The color-spin wave function for $X(3872)$, which is a $I^G(J^{PC})=0^+(1^{++}) $ state, is composed of two states.  Using the quark-antiquark basis $(c\bar{c}) \otimes (q \bar{q})$ with $c,q$ being the charm and light quarks, the two states are $(1 \otimes 1)_{C=1} (V \otimes V)_{J=1}$ and $(8 \otimes 8)_{C=1} (V \otimes V)_{J=1}$,
where the first brackets denote two color singlets $(1)$ or octets $(8)$ combined into color singlet and the second two spin 1 $(V)$ into spin 1.
After subtracting the $K$ factors for the lowest $D \bar{D}^*$ threshold, the color-spin interaction factor is as follows.

$ K_{X(3872)}-K_D-K_{\bar{D}^*}=$ 
\begin{align}
\left(
\begin{array}{cc}
\frac{16}{3 m_c^2}+\frac{16}{3 m_q^2}+\frac{32}{3m_q m_c}   & 0 \\
0 &  - \frac{2}{3 m_c^2}-\frac{2}{3 m_q^2}-\frac{4}{3m_q m_c} 
\end{array}\right), \nonumber
\end{align}
where the factor $\frac{32}{3m_q m_c}$ are due to the subtracted threshold.   One notes that the second diagonal component coming from the color octet element gives an attractive contribution.  However, noting that the $K$ factor coming  from the delta nucleon mass difference is $\frac{16}{m_q^2}$, which  gives a mass difference of around 290 MeV, we find that using a contitutent quark mass of $m_c=1500$ MeV and $m_q=300$ MeV, the attraction can only be around 17.4 MeV assuming that all quarks occupy the size of a nucleon.    
The additional kinetic energy needed to bring the two mesons to 0.3 fm separation, which is the expected typical size of a compact heavy tetraquark configuration, is larger than 200 MeV.  Hence, the attraction is not strong enough to confine the  $X(3872)$ into a compact multiquark configuration. 

A detailed quark model calculation introduces some mixing from the singlet contribution due to the different spatial distributions of heavy and light quarks. This  could introduce  some contribution from the color-color type of interaction and mixing from the color singlet contribution. 
However, it should be noted that at very short distance the color-color interaction is dominated by the Coulomb potential so that it is repulsive for the color octet heavy quark-antiquark pair and do not introduce any additional attraction. 
One can also convert the previous  basis into basis involving $D \bar{D}^*$ color configurations. Then the color octet basis is composed dominantly of color singlet $D$ and color singlet $ \bar{D}^*$ basis: the singlet-singlet to octet-octet fraction is 8 to 1. 
This result furthermore implies that 
the attraction between $D \bar{D}^*$ is small at short distance.

On the other hand, assuming that $J^P=1^+$ for the $T_{cc}$, a better way to describe the color-spin wave function is using the diquark-antidiquark $(ud) \otimes (\bar{c} \bar{c})$ basis, for which the two states are 
$(\bar{3} \otimes 3)_{C=1} (S \otimes V)_{J=1} $ and $(6 \otimes \bar{6})_{C=1} (V \otimes S)_{J=1} $, where the first and second brackets show the color combination and spin with $S$ denoting the spin zero scalar state, respectively.  Using these basis, the $K$ factor for the $T_{cc}$ after subtracting the $K$ factor for the threshold is given as follows.

$K_{T_{cc}}-K_D-K_{D^*}=$ 
\begin{align}
\left(
\begin{array}{cc}
-\frac{8}{ m_q^2}+\frac{8}{3 m_c^2}+\frac{32}{3m_q m_c}   &  \frac{8\sqrt{2}}{m_c m_q} \\
 \frac{8\sqrt{2}}{m_c m_q} &  - \frac{4}{3 m_q^2}+\frac{4}{ m_c^2}+\frac{32}{3m_q m_c} 
\end{array}\right). \nonumber 
\end{align}

\begin{figure*}[bt]
\centering
\begin{tabular}{cc}
\includegraphics[width=.99\columnwidth]{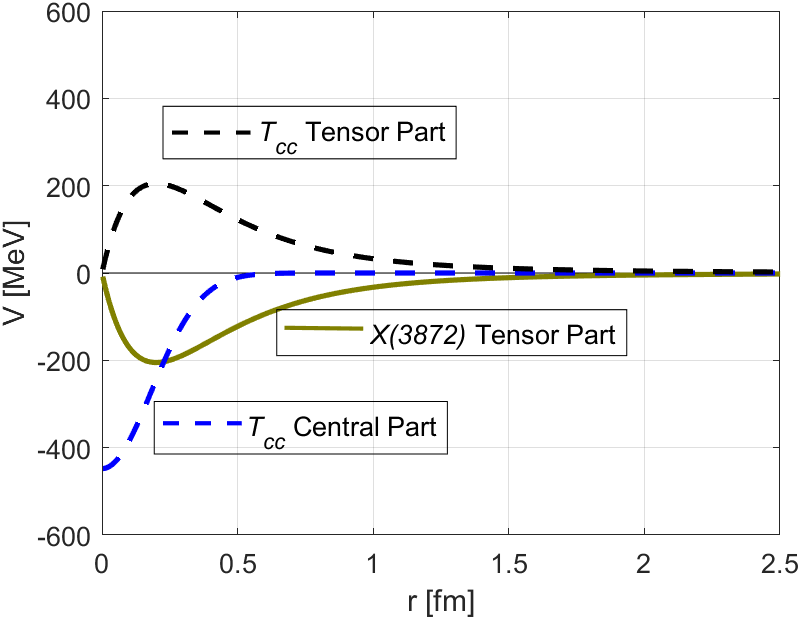}&\includegraphics[width=.99\columnwidth]{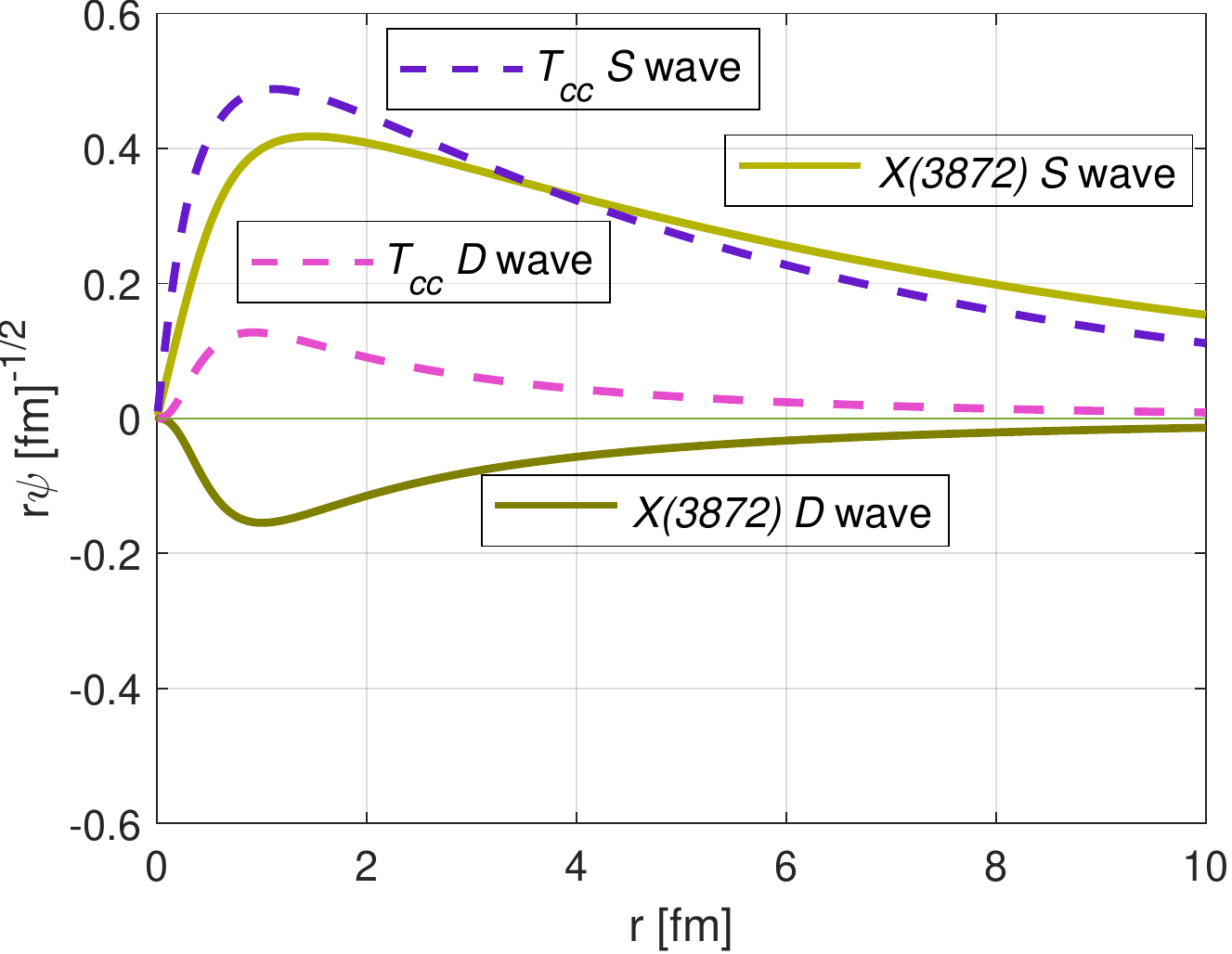}
\end{tabular}
\caption{Left - Tensor potential due to pion exchange contributing to the  $X(3872)$ and $T_{cc}$ with additional short range potential for $T_{cc}$ as a function of distance between two constituents. Right - Wave functions of $X(3872)$ and $T_{cc}$. The wave functions of the deuteron are similar to those of the $T_{cc}$.  }
\end{figure*}

Using the same parameters as above, we find the attraction from the first diagonal component to be around  104.4 MeV, much larger compared to 17.4 MeV for the $X(3872)$.
Diagonalizing the matrix, the attraction is expected to be larger. 
Futhermore, the color-color interaction is expected to produce further attraction in the color triplet or antitriplet quark pair.
This is so because $\lambda^c_i \lambda^c_j=-\frac{16}{3}, -\frac{8}{3} $ and $\frac{2}{3}$ for the $i,j$ quark or antiquark pair in color singlet, triplet(or antitriplet) and octet quark pairs, respectively.
This attraction, coming from the most attractive diquark, was the original motivation for the possible compact $T_{cc}$ configuration\cite{Zouzou:1986qh}. 
Detailed constituent quark model calculations agree on the deeply bound compact configurations for the $T_{bb}$, but do not for the  $T_{cc}$~\cite{Park:2013fda,Park:2018wjk,Noh:2021lqs}.  

At the same time, by analyzing the energy of a compact $T_{cc}$ configuration, one can estimate the attraction between $D-D^*$ at relative distance estimated within that configuration.
Using a simple Gaussian wave function, one finds that at $r_{D-D^*}=0.3$ fm, the attraction is around 200 MeV. This is similar to what has been estimated in the lattice gauge calculation\cite{Ikeda:2013vwa}.   
Therefore, we expect that in the hadronic picture, while there is no short range attraction in the $X(3872)$ channel, there will be an  additional short distance attraction in the $T_{cc}$ channel.

\begin{table*}[t]
\centering
\begin{tabular}{|c|c|c|c|c|c|c|c|}
\hline\hline
Molecule &  Constituents  &  $V_{\text{short}}$ &  $V_\pi$ ($S$ wave) &   $V_\pi$ ($D$ wave-mixing) & $\langle r \rangle$ (fm) & $E_b$ (MeV)  \\
\hline \hline
$d$ & $pn$ &repulsive & attractive & attractive & $ 1.9$   & 2.2  \\ \hline
$X(3872)$ & $D \bar{D^{*}}$ & negligible & negligible & attractive & 3.0   & 0.33  \\ 
\hline
$T_{cc}$ & $D D^*$ & attractive & negligible & attractive & 2.2  & 0.65  \\ 
\hline \hline
\end{tabular}
\caption{Comparison of molecular structures. }
\label{molecular_structure}
\end{table*}

\section{D-wave mixing}

To estimate the D-wave mixing, We will use the pion exchange type of potential between $D-\bar{D^*}$ and $D-D^*$ for the $X(3872)$ and $T_{cc}$ quantum numbers, respectively, that were used in Ref.\cite{Tornqvist:1993ng,Tornqvist:2004qy}.   
\begin{align}
V(r) = -\gamma V_0\Bigg[ \begin{pmatrix}0&-\sqrt{2}\\-\sqrt{2}&1 \end{pmatrix}T_{\pi}(r) \Bigg], \label{potential}
\end{align}
where $V_0 = 1.3$ MeV as determined by the $\pi N$ coupling constant and the pion decay constant, and $\gamma = 3$ and $-3$ for $X(3872)$ and  and $T_{cc}$ channels, respectively\cite{Tornqvist:1993ng}.  
For the $X(3872)$, the attractive tensor potential and the D-wave mixing coming from the off diagonal component in the tensor potential lead  to a loosely bound molecular configuration \cite{Tornqvist:1993ng}. While there is a single pion exchange in  \cite{Tornqvist:1993ng}, this contribution can be neglected. 
On the other hand, because the tensor part is repulsive for the  $T_{cc}$, it  is not bound despite of the common D-wave mixing coming from Eq.~\eqref{potential} which lowers the ground state energy\cite{Tornqvist:1993ng,Tornqvist:2004qy}.

However, as discussed before, one should note that  there 
is a strong attraction at short distance in the $T_{cc}$ channel which will be modeled as follows:    
$V_{\text{short}}(r) = V_{\omega}e^{-\mu_{\omega}^2r^2}$, where $\mu_{\omega}^2 = m_\omega^2 - (m_V - m_P)^2$ is taken to reproduce the $r$ dependence in the lattice result \cite{Ikeda:2013vwa}, which can be viewed as the effect from omega meson exchange.  Furthermore, the overall constant $V_{\omega}=-$448 MeV is taken to reproduce the attraction extracted from the constituent quark model discussed previously. 

This short range potential in addition to the tensor potential (cf. Fig. 1) leads to a loosely bound state of $T_{cc}$  in the molecular configuration with 
binding energy $E_b$ and radius $\langle r \rangle$
given in Table \ref{molecular_structure}.
As discussed before, short range attraction obtained in the quark model is negligible in the $X(3872)$ channel, so that we do not add the additional attraction in that channel.  In fact, even if we add the relatively small attraction there, the result does not change much. 
Their corresponding wave functions are shown in Fig. 1. While the short range attraction is crucial in providing the additional attraction, because the main attraction is coming from the D-wave mixing, the wave function is extended to large distances due to angular momentum barrier so that changing the form or attraction of the potential below 0.3 fm do not give rise to any significant changes in the molecular configuration.

When the formalism is applied to the deuteron case, in which case there is an additional S-wave central potential, one finds $\gamma = 25/3$, and the off diagonal and the lower diagonal components are replaced by $\sqrt{8}$ and $-2$
in the second term of  Eq.~\eqref{potential}, respectively. Again, the strong attraction coming from the D-wave mixing leads to the well known deuteron wave function with a small  binding and a large radius (cf. Table \ref{molecular_structure})  
\cite{Tornqvist:1993ng}.

Therefore, we find that the molecular structures of the $X(3872)$ and $T_{cc}$ composed of heavy mesons are similar to that of the deuteron composed of the nucleons.

\section{Transverse momentum distribution of $d$ and $^3$He: }

Since both the molecular tetraquarks and the deuteron are weakly bound broad molecular states, their transverse momentum dependence in heavy ion collisions should be determined by the distributions of   their respective constituents at the kinetic freeze-out point.  This is so because for widely separated molecules, their cross section with other hadrons should be just the sum of that for the individual constituents.  Such descriptions can be well encoded in the coalescence model, where one could use the observed distribution of the constituents.    

On the other hand, the total yield  could be determined  earlier, 
where the number of constituents are smaller than that at the kinetic freeze-out point where the contributions of excited states are added through feed-down.  

To check our idea and to determine the number of constituents at the formation point, which involves subtracting the contributions of the feed-downs at the formation time,  we will try to fit the transverse momentum distributions as well as the magnitudes of both the deuteron and $^3$He simultaneously.    

For that purpose, we will use a two dimensional coalescence model with a Wigner function of the Gaussian type scaled by $\sigma$ and the coalescence volume $A$.
We calculate  the deuteron distribution  in Pb--Pb collisions at $\sqrt{s_{NN}}=2.76$ TeV \cite{ALICE:2015wav}. 
Along the line, it is necessary to know the transverse momentum distribution of constituents,  the $\sigma$, $A$ and $R_b$. $\sigma$ is a parameter that can be determined by the size of the produced hadron, and the $p_T$ distribution of constituents were fitted using the $p_T$ distribution measured by ALICE collaboration and parametrized as in the Appendix. $A$ is the kinetic freeze-out area. 
Finally, $R_b$ is the ratio between the bare protons and the final protons at the formation point.
$(A,R_b)$ can be determined by fitting the coalescence model results to both the deuteron and $^3$He data.  
This is possible because in the coalescence model, the deuteron production is proportional to $R_b^2/A$ while the  $^3$He production is proportional to $R_b^3/A^2$.  

For the deuteron, we compare the coalescence model result with the  $p_T$ distribution measured in  Pb--Pb collisions at $\sqrt{s_{NN}}=2.76$ TeV and  0--10\% centrality by ALICE collaboration.  For $^3$He, we compare the data at 0-20\% by scaling $A$ according to the charged particle multiplicity.  The proton data were taken at their respective centralities.

\begin{figure}[h]
\begin{center}
\includegraphics[width=.9\columnwidth]{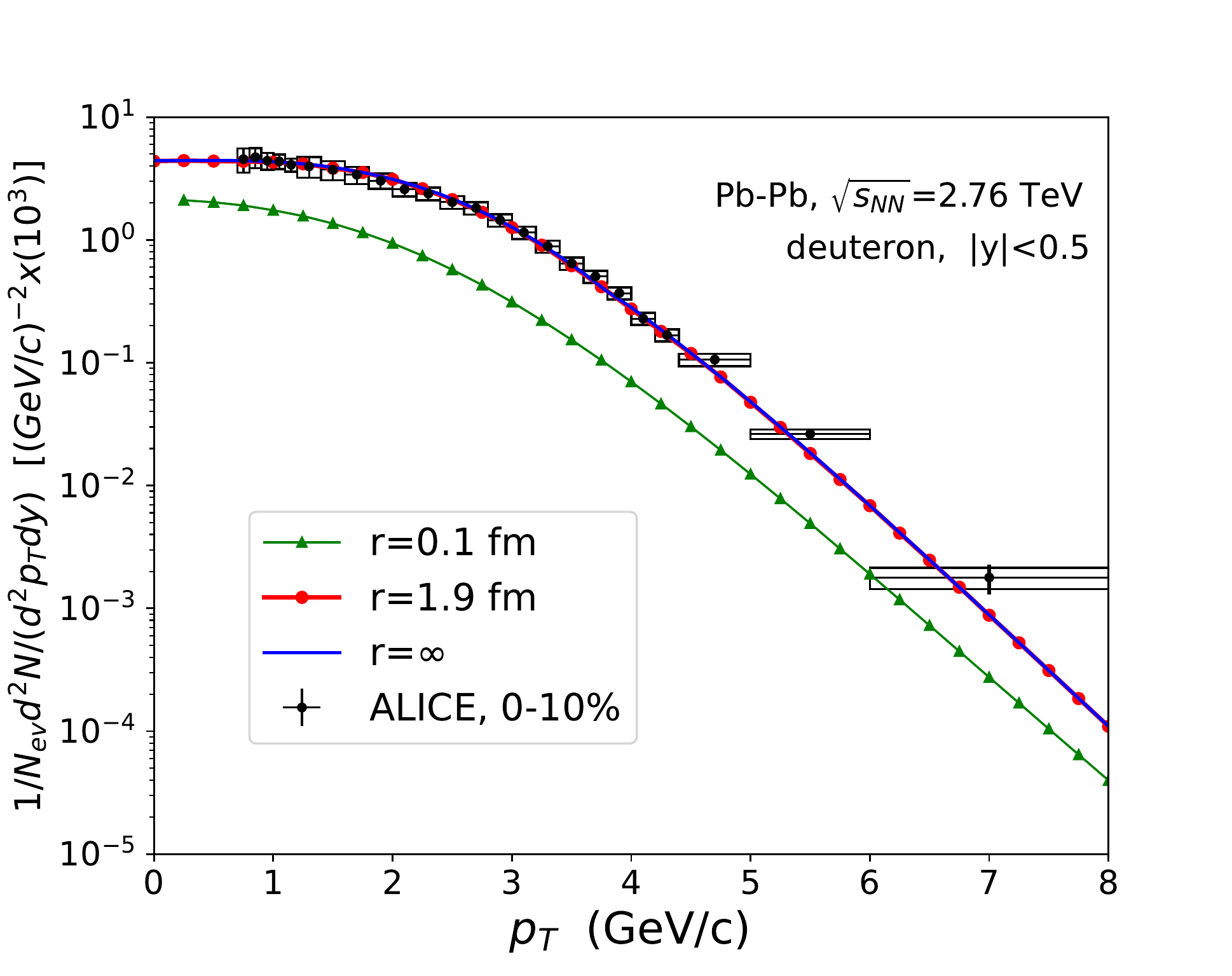}
\includegraphics[width=.9\columnwidth]{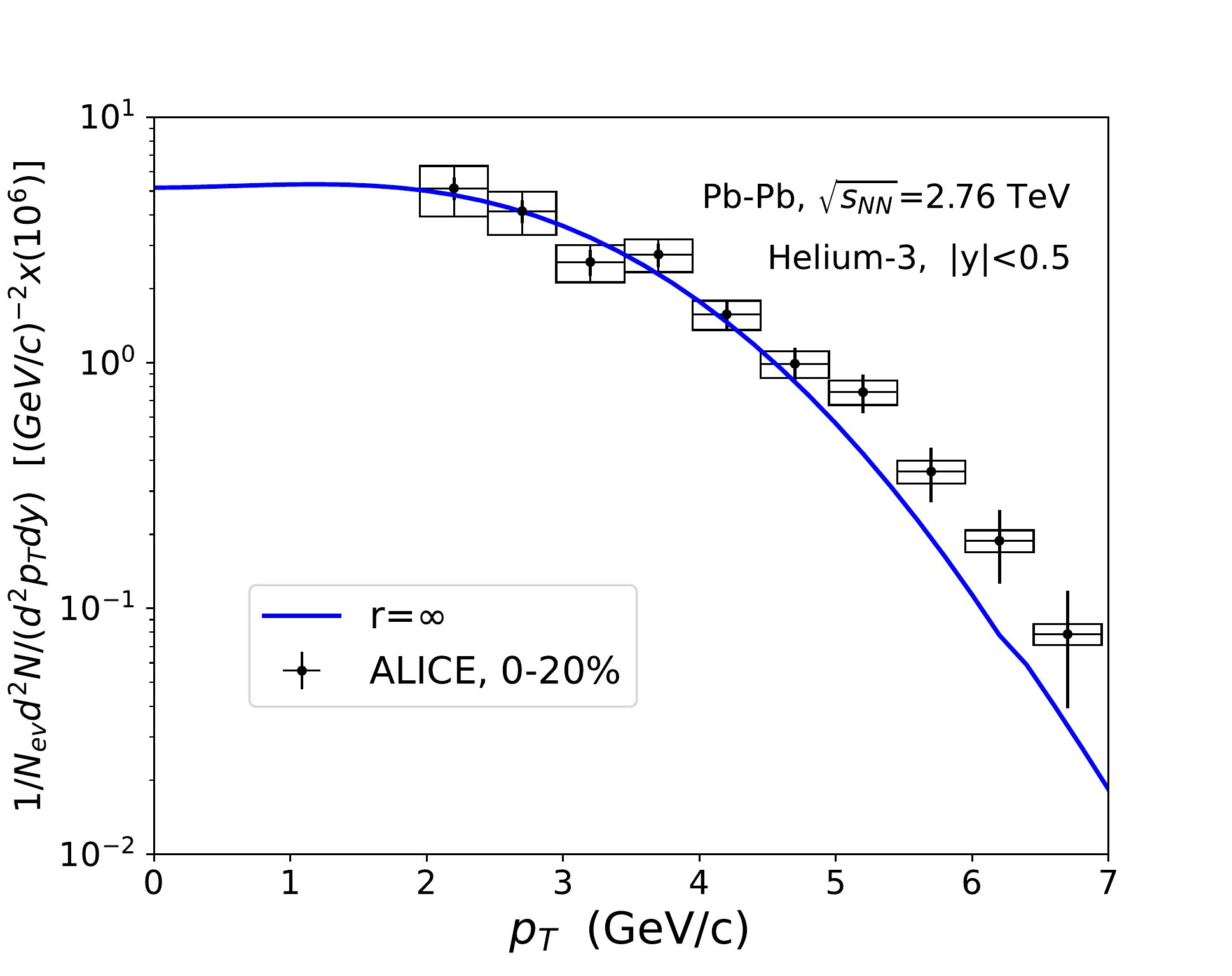}
\end{center}
 \caption{The transverse momentum distributions of deuteron \cite{ALICE:2015wav,ALICE:2017nuf} (top figure) and $^3$He \cite{ALICE:2015wav} (bottom figure) in Pb-Pb collisions at 2.76 TeV and the corresponding fits. }
  \label{fig:profile}
\end{figure}

In the upper graph of Fig. \ref{fig:profile}, we show the transverse momentum distribution for deuteron at 0--10\% centrality using both the original two-dimensional coalescence model \eqref{coal2} with different sizes and $\sigma = \infty$ model \eqref{coal0}. In this calculation we use $A=608 ~\text{fm}^2$, $R_b=0.368$, proton mass 938 MeV and neutron mass 939 MeV. Also, since $\sigma = \sqrt{8/3} r$ Ref.~\cite{Sun:2018mqq}  and the radius of deuteron is taken to be $r_d=1.9$,  we used $\sigma=3.1$ fm. The 0--10\% proton distribution was obtained using the average of the 0--5\% distribution and the 5--10\% distribution. 
As can be seen in the upper graph of Fig. \ref{fig:profile}, the coalescence formula with the physical radius $r_d=1.9 $ fm and that obtained 
in the $\sigma \rightarrow \infty$ limit lies almost on top of each other and well  reproduces the experimental results.  The effect of small radius becomes visible only when $\sigma << 1/\sqrt{mT}$ ($m$ is the reduced mass of its constituent) such as $r=0.1$ fm shown in the figure. 
In the lower graph of Fig. \ref{fig:profile}, the $^3$He data are well reproduced with the same $R_b$, although at higher $p_T$ the model lies at the lower boundary of the error bar.  

Therefore, since the expected radius in the molecular configurations of both the  $X(3872)$ and $T_{cc}$ are larger than that of the deuteron, we can calculate the transverse momentum dependence for these molecular tetraquark configuration using $\sigma \rightarrow \infty$.

It is interesting to note that the factor $R_b $ obtained to simultaneously fit the deuteron and $^3$He are close to what is expected of the fraction of bare proton number using statistical hadronization model(SHM)\cite{Hwa:2004yg} at the chemical freeze-out point. Therefore, we will also use the SHM with charm quarks to estimate the corresponding $R_b$ factor for the $D$ and $D^*$ mesons that will be used as the input in the coalescence model for $X(3872)$ and $T_{cc}$.

\section{Transverse momentum distribution of $X(3872)$ and $T_{cc}$:}

\begin{figure}[h]
\centering
{
\includegraphics[width=1.05\columnwidth]{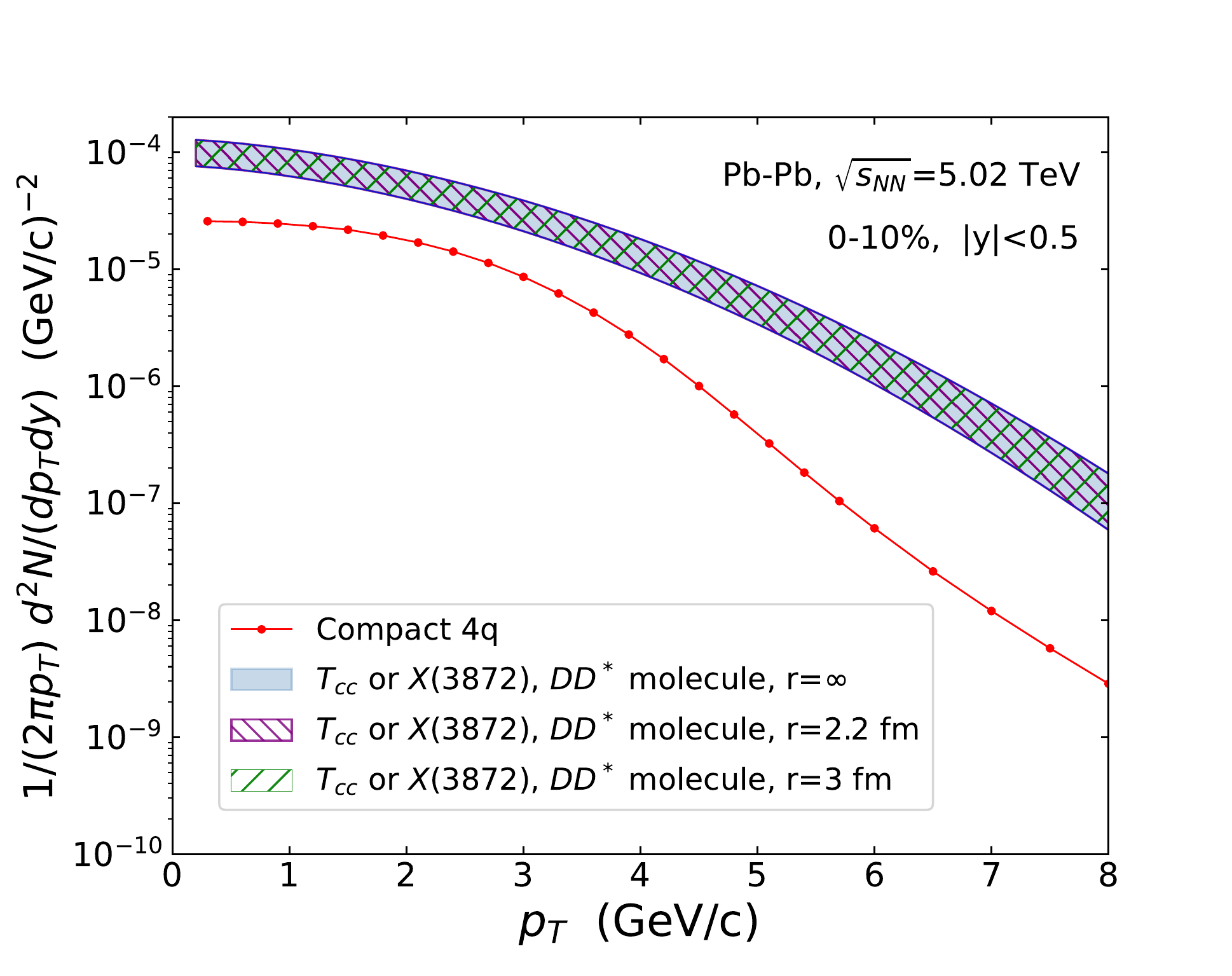}
}
\caption{The transverse momentum distribution of $X(3872)$ and $T_{cc}$ (0--10\% event). The bands for molecular configurations are due to the uncertainties in the input data.} \label{XT}
\end{figure}

Fig. \ref{XT} shows coalescence model result for the transverse momentum distribution of $T_{cc}$ and $X(3872)$ at  centrality 0--10\% for Pb--Pb collisions at $\sqrt{s_{NN}}=5.02$ TeV. 
Assuming a molecular configuration of  $X(3872)$, its production will be determined by the distribution of 
$D^0$ and $\bar{D}^{*0}$ at the kinetic freeze-out point.  
We fitted the observed $D^0$ and $\bar{D}^{*+}$ data in Pb--Pb collisions at $\sqrt{s_{NN}}=5.02$ TeV.
As the the $D^{*0}$ $p_T$ distribution is not currently measured, we used the measured  $\bar{D}^{*+}$ distribution here.  Taking into account feed downs, we expect that  31$\%$ of the measured $D^{0}$ will participate in the coalescence (See Appendix for details).  
Yield of $D^0$ in Pb--Pb collisions at 5.02 TeV is $dN_{D^0}/dy$ = 6.819 \cite{ALICE:2021rxa}, so that the expected bare number is 2.12.

Since the chemical potential can be almost neglected, a molecular configuration of $T_{cc}$ will have almost identical transverse momentum distribution of $X(3872)$ as given in Fig. \ref{XT}.  
The  band in Fig. \ref{XT} shows the statistical uncertainty of the input $D,D^*$ data transferred  into the coalescence result.
One also notes that for the radii given in Table \ref{molecular_structure}, the results are almost identical to the result given in the $\sigma \rightarrow \infty$ limit.

We also plot, the expected transverse momentum distribution of $T_{cc}$ as given in Ref. \cite{Cho:2019syk}, obtained assuming that it has a compact multiquark configuration.  The  result in Ref. \cite{Cho:2019syk} given for collision energy at 2.76 TeV, has been multiplied by  1.63 to take into account the difference in the charm quark number squared at 5.02 TeV as given in \cite{ExHIC:2017smd} to compare with the result for the molecular configuration calculated here.  As can be seen in the figure, 
the expected transverse momentum distributions for two possible configurations of the $T_{cc}$ are markedly different in addition to the difference in the total yields.  
Therefore, 
once the transverse momentum distribution of $T_{cc}$ is measure, we will be able to discriminate its structure.

\section{Total yields and summary}

One of us has previously noted that for an exotic hadron to be compact, it has to include multiple heavy quarks and that it could therefore be abundantly produced in a relativistic heavy ion collision, where many heavy quark-antiquark pairs are produced  \cite{Lee:2007tn,Proceedings:2007ctk}. Later, two of us with the ExHIC collaboration have shown that one could discriminate whether an exotic is a molecular configuration or a compact multiquark state, from its production rate in a heavy ion collision  \cite{ExHIC:2017smd,ExHIC:2010gcb,ExHIC:2011say}. Recently,  the CMS collaboration at CERN measured the $X(3872)$ for the first time in a heavy ion collision, and found its production ratio to $\psi(2S)$ to be anomalously larger than that in pp collision \cite{CMS:2021znk}. According to the ExHIC papers, enhanced production of $X(3872)$ in heavy ion collisions implies that  its structure is a loosely bound molecular configuration.  Therefore, we 
estimate the total yields of hadrons per unit rapidity at central rapidity in a collision event by integrating the  calculated transverse momentum distributions in the $\sigma \rightarrow \infty$ limit.

Table \ref{tot-yield-nuclei} shows the ratio between the total yield of $d$ and $^3$He calculated using the coalescence model and the statistical model with $T_H=156$ MeV and $\mu_B =0.7$ MeV. 
The coalescence model reproduces the SHM result for the deuteron yield but slightly underestimates the corresponding  value for  $^3$He.  

The total yields for the tetraquark states are given in Table \ref{tot-yield-charm}.
Assuming the tetraquark states are of molecular configurations, the coalescence model results, which are the same for the $X(3872)$ and the $T_{cc}$, are appreciably larger than that from the statistical model result with charm (SHMc) \cite{Andronic:2021erx}: the ratio is 2.47. It should be noted that unlike the deuteron case where one has to consider the feeddown contribution for both constituents, one only eliminates those for $D$ and not for $D^*$ or $\bar{D}^*$.   However, assuming that the $T_{cc}$ is composed of a compact multiquark configuration, the coalescence model gives a smaller value than that expected from the SHMc. 
The last colume in Table \ref{tot-yield-charm} shows the yield ratio of tetraquark states to $\psi(2S)$ in the SHMc.  As can be seen in the table, the ratio is $0.806 \pm 0.234$ if the tetraquark is assumed to be of molecular configuration.  
Although the experimental measurement is at high transverse momentum and that the data takes into account the different branching ratios into $J/\psi \pi \pi $, considering further suppression of $\psi(2S)$ in heavy ion collision, the observed ratio seems consistent with a molecular picture for $X(3872)$.

\begin{table}[h]
\centering
\begin{tabular}{c|c|c}
\hline\hline
Nucleus &  $N^{Nucleus}_{SHM}/N^{p}_{SHM}$ & $N^{Nucleus}_{coal}/N^{p}_{SHM}$   \\
\hline \hline
$d$ & $9.07 \times 10^{-3}$  & $8.84 \times 10^{-3} $ \\ \hline
$^3$He & $2.68 \times 10^{-5} $ & $ 2.03 \times 10^{-5} $ \\ 
\hline \hline
\end{tabular}
\caption{The yield ratio of light nucleus with proton  in Pb--Pb collisions at $\sqrt{s_{NN}}=2.76$ TeV. For deuteron and $^3$He the centralities are 0--10 \% and 0--20 \%, respectively.}
\label{tot-yield-nuclei}
\end{table}

\begin{widetext}
\begin{table}[h]
\centering
\begin{tabular}{c|c|c|c}
\hline\hline
Tetraquark &$dN_{coal}/dy$ & $N_{coal}/N^{X(3872)}_{SHMc}$   
& $N_{coal}/N^{\psi(2S)}_{SHMc} $
\\
\hline \hline
$DD^*$ molecule & $ (2.45 \pm 0.71 )  \times 10^{-3} $  & $ 2.47 \pm 0.716 $ &  $0.806 \pm 0.234$ \\ \hline
Compact \ 4$q$  &  6.2 $\times 10^{-4}$  & $ 6.25 \times 10^{-1} $ &  0.204 \\
\hline \hline 
\end{tabular}
\caption{The first column shows the total yield of the tetraquark depending on its structure calculated by the coalescence model in Pb--Pb collisions at $\sqrt{s_{NN}}=5.02$ TeV at 0--10\% centrality.  The remaining columns show their ratios to the statistical hadronization model with charm (SHMc) \cite{Andronic:2021erx}.    Here we used $dN_{\psi(2S)}/dy = 3.04 \times 10^{-3}$  and $N_{X(3872)}/N_{\psi(2S)} = 0.326$ obtained in SHMc.}
\label{tot-yield-charm}
\end{table}
\end{widetext}

We have argued why $X(3872)$ should be a molecular configuration while $T_{cc}$ could be either a molecular configuration or a compact mutlqiaruk state.
Our analysis suggest that the anomalously large yield ratio $X(3872)/\psi(2S)$ in heavy ion collision indeed seems consistent with the moelcular picture of $X(3872)$.  However, as the present experimental result is at higher transverse momentum than what we calculated here, further measurements of these as well as the newly observed exotics \cite{Gershon:2022xnn} in heavy ion collision will be able to clearly discriminate a compact multiquark configuration and a loosely bound molecular configuration.

\newpage

\appendix

\section{$\pi$-exchange potential}

We use the following  tensor potential as given in Ref. \cite{Tornqvist:1993ng,Tornqvist:2004qy}.   

\begin{flalign}
T_{\pi}(r) & = \frac{\mu^2r^2+3\mu r + 3}{m_\pi^3r^3}e^{-\mu r} - \frac{\Lambda^2 r^2 + 3\Lambda r + 3}{m_\pi^3r^3}e^{-\Lambda r}&&\\
&\quad - \frac{(\Lambda^2 - \mu^2)(\Lambda r +1)}{2m_\pi^3r}e^{-\Lambda r}, \nonumber
\end{flalign}
where  $\Lambda=$1300 MeV and 
$\mu^2 = m_{\pi}^2 - (m_V - m_P)^2$ stemming from the recoil effect between pseudoscalar and vector mesons.

\section{Coalescence model}

We use the following two particle coalescence model in two-transverse dimension at central rapidity, which for the deuteron combines two nucleons.
\begin{align}
\frac{d^2 N_d}{d^2 P_T} = g_d & \int d^2 x_1 d^2 x_2 d^2 p_{1T} d^2 p_{2T}  \frac{d^2 N_p}{dA_L d^2 p_{1T}} \frac{d^2 N_n}{dA_L d^2 p_{2T}} 
\nonumber \\
& \times W_d(\vec{r}, \vec{k})  \delta^{(2)} \left( \vec{P}_{T} - \vec{p}_{1T} - \vec{p}_{2T} \right), \label{coal2}
\end{align} 
where $g_d$, $A_L$ and $W_d(\vec{r}, \vec{k}) $ are the deuteron statistical factor divided by those of the constituents, the kinetic freeze-out area in the Lab frame and the Winger function of deuteron in the CM frame, respectively. Here, we use a two-dimensional Gaussian-type Wigner function given as follows:
\begin{align}
W_d(\vec{r}, \vec{k}) = 4 \exp{\left( -\frac{r_1'^2}{\sigma^2} -\sigma^2 k_1'^2 \right) }. \label{Wig-func}
\end{align}

As emphasised in  \cite{Cho:2019lxb}, to properly take into account deuterons with large transverse momenta,  the  Wigner function should be defined in the deuteron rest frame with the following relative distance and momentum: 
\begin{align}
\vec{R}^{\, \prime} = \frac{m_1 \vec{x}^{\, \prime}_{1T} + m_2 \vec{x}^{\, \prime}_{2T}}{m_1 + m_2}, \quad  \vec{r}^{\, \prime}_1 = \vec{x}^{\, \prime}_{1T} - \vec{x}^{\, \prime}_{2T}, 
\nonumber \\
 \vec{k}^{\, \prime}= \vec{p}^{\, \prime}_{1T} + \vec{p}^{\, \prime}_{2T}, \quad \vec{k}^{\, \prime}_1 = \frac{m_2 \vec{p}^{\, \prime}_{1T} - m_1 \vec{p}^{\, \prime}_{2T} }{m_1 + m_2}. 
\end{align}
The parameter $\sigma$ in Eq.~\eqref{Wig-func} is related to the deuteron radius by $\sigma =  \sqrt{8/3} r_d$. 
Assuming a spatially homogeneous $\frac{d^2 N_p}{dA_L d^2 p_{1T}}$, the transverse momentum distribution reduces as follows:
\begin{align}
\frac{d^2 N_d}{d^2 P_T} = \frac{g_d}{A} & (2\sqrt{\pi})^2 \sigma^2  \int d^2 p_{1T} d^2 p_{2T}  \frac{d^2 N_p}{d^2 p_{1T}} \frac{d^2 N_n}{d^2 p_{2T}}  
\nonumber \\
& \times  \exp{ (-\sigma^2 k_1'^2) } \delta^{(2)} \left( \vec{P}_{T} - \vec{p}_{1T} - \vec{p}_{2T} \right).    \label{d_distribution}
\end{align} 

In the $\sigma \rightarrow \infty$ limit,
\begin{align}
\frac{d^2 N_d}{d^2 P_T} 
&=  g_{d}  (2\pi)^2 \left( \frac{\gamma}{A} \right)  \frac{d^2 N_{p}}{d^2p_{pT}}|_{\vec{p}_{pT}= \frac{m_p}{M} \vec{P}_T} \frac{d^2 N_{n}}{d^2p_{nT}}|_{ \vec{p}_{nT}= \frac{m_n}{M} \vec{P}_T} , \label{coal0}
\end{align}

For the deuteron distribution  in Pb--Pb collisions   at  in $\sqrt{s_{NN}}=2.76$ TeV with 0--10\% centrality, we use  $A=608~\text{fm}^2$.  
For $^3$He, we compare our calculation to the data by ALICE collaboration measured for 0--20\% centrality \cite{ALICE:2015wav}.  0--20\% proton distribution was obtained by averaging the 0--5\%, 5--10\% and 10--20\% distributions. Because we have a different centrality, the kinetic freeze-out area $A$ that we used in the deuteron fit will be rescaled by the ratio of the charged particle multiplicities between different centralities.  This gives $A = 507$ fm$^2$.  
  
To estimate the feed down contribution, we use the SHM \cite{Hwa:2004yg}
\begin{align}
N^{stat}_h &= V_H \frac{g_h}{2\pi^2} \int^{\infty}_0 \frac{p^2 dp}{\gamma^{-1}_h e^{E_h / T_H} \pm 1} ,
\nonumber \\
&\approx \frac{\gamma_h g_h V_H}{2\pi^2} m_h^2 T_H K_2 \left( \frac{m_h}{T_H} \right) ,
\end{align}
where $g_h$ is the degeneracy factor of hadron, $\gamma_h$ is fugacity, $K_2$ is the modified Bessel function of the second kind, and $V_H$ and $T_H=156$ MeV are the hadronization volume and temperature at chemical freeze-out. 

For the calculation of molecular configuration of $X(3872)$ and $T_{cc}$, we present the result 
for Pb--Pb collisions at $\sqrt{s_{NN}}=5.02$ TeV with 0--10\% centrality. The coalescence area $A$ in Pb--Pb collisions at $\sqrt{s_{NN}}=5.02$ \ TeV was rescaled according to the charged particle multiplicity: using $dN/d\eta^{\text{Pb-Pb}, 2.76 \ \text{TeV}}_{0-10\%} = 1447.5 \pm 39$ \cite{Alice:PbPb_2.76} and $dN/d\eta^{\text{Pb-Pb}, 5.02 \ \text{TeV}}_{0-10\%} = 1764 \pm 25$ \cite{Alice:PbPb_5.02}, $A$ was taken to be $A_{0-10\%}^{5.02 \ \text{TeV}}=$ 740.5 fm$^2$. 
We fitted the observed $D^0$ and $\bar{D}^{*+}$ data in Pb--Pb collisions at $\sqrt{s_{NN}}=5.02$ TeV.  
As the $D^{*0}$ $p_T$ distribution is not currently measured, we used the measured $\bar{D}^{*+ }$ instead.   We again use the statistical model to estimate the fraction of the $D^0$ participating in coalescence.
The decay channels $D^{*}(2007)^{0} \rightarrow D^{0} \pi^{0}$ (BR = 64.7$\%$), $D^{*}(2007)^{0} \rightarrow D^{0} \gamma$ (BR = 35.3$\%$), and $D^{*}(2010)^{+} \rightarrow D^{0} \pi^{+}$ (BR = 67.7$\%$) are considered.
We then find that  31$\%$ of the measured $D^{0}$ participates in the coalescence.

\section{Parameterized transverse momentum distributions}
\label{fit}

Here, we provide the distributions of the constituents needed in the coalescence calculations, which were obtained by  fitting the transverse 
momentum distribution measured in experiments. An exponential function and a 
power-law type function were used for low and high $p_T$, respectively:
 \begin{align}
\frac{d N}{d^2 p_Tdy} &= a e^{-b(p_T/p_1)^c} \quad \quad\qquad (\ p_T < p_c), 
\nonumber \\
&= \frac{d}{[1 + (p_T/p_0)^2]^e} \quad ( \ p_T > p_c), 
\label{c1}
\end{align}
where $a, \ b, \ c, \ d, \ e$ and $p_0$ are fitting parameters. Here, $p_1 = 1.0$ GeV and $|y| < 0.5 $.  The values of $p_c$ and the fitting parameters are shown in Table \ref{param}.  
Our form in Eq.~\eqref{c1} is close to the Blast wave form\cite{Alice:PbPb_2.76} but better reproduces the proton data for $p_T>3$ GeV.

\begin{widetext}

\begin{table}[h]
\centering
\begin{tabular}{c|c|c|c|c|c|c|c|c|c}
\hline\hline
 & $\sqrt{s_{NN}}$  (GeV)  & $\%$ & a (GeV)$^{-2}$ & b & c & d (GeV)$^{-2}$ & e & $p_0$ (GeV) & $p_c$ (GeV)  \\
\hline \hline
$p$ & 2.76 & 0--5 & 4.551 & 0.376& 2.282& 6.670& 7.995& 3.588 &2.0\\ \hline
$p$ & 2.76 & 5--10 & 3.727 & 0.3477& 2.42& 7.384& 5.55& 2.644 &2.0\\ \hline
$p$ & 2.76 & 10--20 & 2.937 & 0.3812& 2.333& 4.415& 6.719& 3.149 &2.0\\ \hline
$D^0_u$ & 5.02 & 0--10 & 0.7537 & 0.3224& 1.788& 3.028$\times10^{8}$& 3.053& 0.08034 &4.5\\
\hline
$D^0_l$ & 5.02 & 0--10 & 0.5885 & 0.3463& 1.836& 9.959$\times10^{12}$& 3.013& 0.01264 &4.0\\ \hline
$D^{*+}_u$ & 5.02 & 0--10 & 0.2673 & 0.41& 1.568& 1.62$\times10^{10}$& 2.824& 0.02545 &4.5\\ \hline
$D^{*+}_l$ & 5.02 & 0--10 & 0.204 & 0.428& 1.55& 4.723$\times10^{8}$& 3.057& 0.06464 &4.5\\ 
\hline \hline
\end{tabular}
\caption{ The fitting parameters. The subscripbs in the first column corresponds to the fits for the  upper (u) and lower (l) bounds in Fig. \ref{XT}, respectively. }
\label{param}
\end{table}

\end{widetext}

In Pb-Pb collisions at $\sqrt{s_{NN}}=2.76$ TeV \cite{Alice:PbPb_2.76}, the transverse momentum distribution of protons was used for deuteron and $^3$He calculations.

In Pb-Pb collisions at $\sqrt{s_{NN}}=5.02$ TeV \cite{ALICE:2021rxa}, 
the transverse momentum distribution of $D$ and  $D^{*}$~meson were used for $X(3872)$ calculations.

\section*{Acknowledgements}
The work was supported by Samsung Science and Technology Foundation under Project Number SSTF-BA1901-04, and by the National Research Foundation of Korea (NRF) grant funded by the Korea government (MSIT) under Project No.2018R1A5A1025563, No. 2019R1A2C1087107, No. 2021R1A2C1009486, and No. 2022R1A2C1011549.

\end{document}